\documentclass[a4paper]{jpconf}
\usepackage{graphicx}
\begin{document}

\def\gsim{\mathop {\vtop {\ialign {##\crcr 
$\hfil \displaystyle {>}\hfil $\crcr \noalign {\kern1pt \nointerlineskip } 
$\,\sim$ \crcr \noalign {\kern1pt}}}}\limits}
\def\lsim{\mathop {\vtop {\ialign {##\crcr 
$\hfil \displaystyle {<}\hfil $\crcr \noalign {\kern1pt \nointerlineskip } 
$\,\,\sim$ \crcr \noalign {\kern1pt}}}}\limits}

\title{Roles of Critical Valence Fluctuations in Ce- and Yb-Based Heavy Fermion Metals}

\author{Shinji Watanabe, Kazumasa Miyake}

\address{Graduate School of Engineering Science, Osaka University, Toyonaka, Osaka 560-8531, Japan}


\begin{abstract}
The roles of critical valence fluctuations of Ce and Yb are discussed as a key origin of 
several anomalies observed in Ce- and Yb-based heavy fermion systems. 
Recent development of the theory has revealed that a magnetic field is an efficient control parameter 
to induce the critical end point of the first-order valence transition. 
Metamagnetism and non-Fermi liquid behavior caused by this mechanism are discussed 
by comparing favorably with CeIrIn$_5$, YbAgCu$_4$, and YbIr$_2$Zn$_{20}$. 
The interplay of the magnetic order and valence fluctuations offers a key concept for understanding 
Ce- and Yb-based systems. It is shown that suppression of the magnetic order 
by enhanced valence fluctuations gives rise to the coincidence of the magnetic-transition point 
and valence-crossover point at absolute zero as a function of pressure or magnetic field. 
The interplay is shown to resolve the outstanding puzzle in CeRhIn$_5$ in a unified way.  
The broader applicability of this newly clarified mechanism is discussed by surveying promising materials 
such as YbAuCu$_4$, $\beta$-YbAlB$_4$, and YbRh$_2$Si$_2$.  
\end{abstract}

\section{Introduction}

In this article, we discuss the roles of critical valence fluctuations of Ce and Yb 
as a key origin of several anomalies observed in the Ce- and Yb-based heavy fermion systems. 
Recent development of the theory and experiments have revealed that 
the valence fluctuations appear ubiquitously by tuning the control parameters of 
the magnetic field and pressure, giving rise to broader consequences 
than previously recognized. 
First, we briefly survey accumulated experiments on the anomalies related to 
the valence transition and its fluctuations, 
and explain how the theoretical understanding has been achieved so far. 
Then, we discuss the recent development of theory and experimental results 
focusing on the prototypical materials. 

\begin{figure}
\begin{center}
\includegraphics[width=150mm]{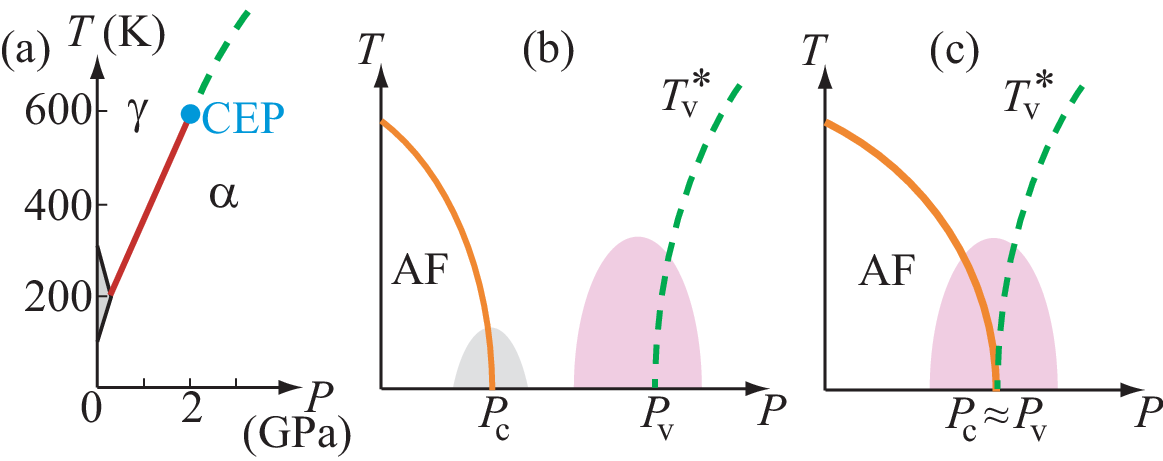}
\end{center}
\caption{\label{fig:PcPv}(color online) 
(a) $T$-$P$ phase diagram of Ce metal. 
The first-order valence transition line (bold line) terminates at a 
critical end point (filled circle). 
Schematic $T$-$P$ phase diagrams for 
(b) $P_{\rm c} < P_{\rm v}$, 
and 
(c) $P_{\rm c} \approx P_{\rm v}$. 
The dashed line represents valence-crossover temperature $T_{\rm v}^{*}$ (see text). 
In (b) and (c), shaded regions represent superconducting phases. 
}
\end{figure}

Valence transition is an isostructural phase transition with a valence of the materials element 
showing a discontinuous jump. 
A typical example is known as the $\gamma$-$\alpha$ transition in Ce metal~\cite{Ce}, where 
the first-order valence transition (FOVT) occurs in the temperature-pressure $(T,P)$ phase diagram 
while keeping the fcc lattice structure, as shown in Fig.~\ref{fig:PcPv}(a). 
The critical end point (CEP) of the FOVT is located at $(T,P)=$(600~K, 2~GPa). 
The valence of Ce changes discontinuously between $\rm Ce^{+3.03}$ ($\gamma$ phase) and $\rm Ce^{+3.14}$ 
($\alpha$ phase) at $T=300$~K~\cite{Wohlleben}. 
YbInCu$_4$ is also well known as a prototypical material for the isostructural FOVT~\cite{Felner,Sarrao1996}, 
where as $T$ decreases, the Yb valence changes discontinuously from $\rm Yb^{+2.97}$ to $\rm Yb^{+2.84}$ 
at $T=42$~K~\cite{Matsuda}. 
As diverging density fluctuations in the liquid-gas transition, valence fluctuations diverge 
at the CEP of the FOVT. 
When the temperature of the CEP is suppressed by controlling the materials parameters 
and enters the Fermi degeneracy regime, diverging valence fluctuations are considered to 
be coupled to the Fermi-surface instability. 
This multiple instability is considered to be a key mechanism for understanding anomalies observed in 
Ce- and Yb-based heavy-fermion systems. 
Such phenomena have been detected in CeCu$_2$Ge$_2$~\cite{Jaccard} and 
CeCu$_2$Si$_2$~\cite{holms}, where 
remarkable enhancement of the superconducting transition temperature appears at the pressure $P_{\rm v}$ 
larger than $P_{\rm c}$ corresponding to the antiferromagnetic (AF) quantum critical point (QCP) 
(see Fig.~\ref{fig:PcPv}(b)), 
and in CeCu$_2$(Si$_{x}$Ge$_{1-x}$)$_2$ where two separate domes of the superconducting phase 
have been found under pressure~\cite{yuan}. 
In these materials, 
near $P=P_{\rm v}$, remarkable anomalies such as the $T$-linear resistivity and enhancement of 
the residual resistivity have been observed, which clearly exhibit distinct behavior of resistivity 
near $P=P_{\rm c}$~\cite{Jaccard,holms,yuan}. 
Furthermore, these materials show that the coefficient $A$ obtained by the $T^2$ fitting to 
the lowest-$T$ regime of the resistivity decreases by 2-3 orders of magnitude 
when $P$ increases 
across $P=P_{\rm v}$.

Theoretically, it has been pointed out that 
enhanced Ce-valence fluctuations are the possible origin of the anomalies near $P=P_{\rm v}$ 
in the CeCu$_2$(Si/Ge)$_2$ systems~\cite{MNO}. 
It has also been pointed out that 
the sudden drop of the coefficient $A$ at $P=P_{\rm v}$ 
can be understood as a sharp change of the valence of Ce on the basis of the Gutzwiller arguments 
in the periodic Anderson model~\cite{OM}. 
It has been also shown that enhanced valence fluctuations cause the $T$-linear resistivity~\cite{holms} 
and enhancement of residual resistivity~\cite{MM}. 
The emergence of the superconducting dome near $P\sim P_{\rm v}$ 
in Fig.~\ref{fig:PcPv}(b) has been explained 
by the valence-fluctuation-mediated superconductivity 
shown by the slave-boson mean-field theory taking account of the Gaussian fluctuations 
in the periodic Anderson model~\cite{OM}. 
The density-matrix renormalization group (DMRG) calculation applied to the same model 
in one spatial dimension has also shown 
the enhancement of the superconducting correlation in the same parameter regime~\cite{WIM}. 
Pairing symmetry of density-fluctuation (i.e., valence-fluctuation)-mediated superconductivity 
has also been analyzed on the basis of the phenomenological model~\cite{ML}.

Furthermore, recent NQR measurement in CeCu$_2$Si$_2$ has detected change of the Cu-NQR frequency 
near $P=P_{\rm v}$, 
suggesting that the Ce valence changes near $P=P_{\rm v}$~\cite{fujiwara}. 
Detailed measurement of the $T$ dependence of the specific heat and the upper critical field 
in CeCu$_2$Si$_2$ under pressure concluded that the pairing symmetries of the superconducting phases 
near $P=P_{\rm c}$ and $P=P_{\rm v}$ are different, suggesting that a pairing mechanism different 
from the antiferromagnetic spin-fluctuation mediated one is realized near $P_{\rm v}$~\cite{Lengyel}.
For a detailed summary of the theory and related experiments up to this stage, readers can refer 
to Ref.~\cite{M07} and also Ref.~\cite{WM_pss}. 

Recent development of the theory of the quantum critical end point (QCEP) of the FOVT
has revealed that valence fluctuations play a key role 
in the other Ce- and Yb-based heavy fermion systems more ubiquitously. 
We discuss the crucial roles of valence fluctuations in the following sections. 
In \S2, we discuss the properties of the valence transition and valence crossover of Ce or Yb 
by comparing theoretical phase diagrams with Ce- and Yb-based systems. 
In \S3, we discuss the fact that the QCEP of the FOVT can be induced rather easily 
by applying the magnetic field to the Ce- and Yb-based systems, which causes various anomalies 
such as metamagnetism and non-Fermi liquid behavior. 
We discuss that this mechanism gives a unified explanation for CeIrIn$_5$, YbAgCu$_4$, and 
YbIr$_2$Zn$_{20}$.  
In \S4, we discuss the interplay of magnetic order and valence fluctuations, giving rise to 
coincidence of the magnetic transition and sharp valence crossover at $T=0$~K, 
as shown in Fig.~\ref{fig:PcPv}(c).  
We show how the interplay resolves the outstanding puzzle in CeRhIn$_5$ and also in YbAuCu$_4$.  
The summary and outlook are given in \S5. 
The broader applicability of the present mechanism is discussed by surveying promising materials 
such as $\beta$-YbAlB$_4$ and YbRh$_2$Si$_2$.

\section{Model and phase diagrams}
\subsection{Minimal model for Ce- and Yb-based heavy fermions exhibiting valence transition} 

We consider the simplest minimal model, which describes an essential part of the physics 
in  Ce- and Yb-based heavy fermion systems exhibiting valence transition, as follows: 
\begin{equation}
{\cal H}=H_{\rm c}+H_{\rm f}+H_{\rm hyb}+H_{U_{\rm fc}}, 
\label{eq:PAM} 
\end{equation}
where 
$H_{\rm c}=\sum_{{\bf k}\sigma}\varepsilon_{\bf k}
c_{{\bf k}\sigma}^{\dagger}c_{{\bf k}\sigma}$ 
represents the conduction band, 
$H_{\rm f}=\varepsilon_{ \rm f}\sum_{i\sigma}n^{ \rm f}_{i\sigma}
+U\sum_{i=1}^{N}n_{i\uparrow}^{ \rm f}n_{i\downarrow}^{ \rm f}$ 
the f level $\varepsilon_{\rm f}$ and onsite Coulomb repulsion $U$ for f electrons, 
$H_{\rm hyb}=V\sum_{i\sigma}\left(
f_{i\sigma}^{\dagger}c_{i\sigma}+c_{i\sigma}^{\dagger}f_{i\sigma}
\right)$ 
the hybridization $V$ between f and conduction electrons, 
and 
$
H_{U_{\rm fc}}=
U_{\rm fc}\sum_{i=1}^{N}n_{i}^{ \rm f}n_{i}^{ c}
$ 
the Coulomb repulsion $U_{\rm fc}$ between f and conduction electrons. 
The $H_{U_{\rm fc}}$ term is a key ingredient for explaining the valence transition 
as well as the various anomalies 
caused by enhanced Ce- or Yb-valence fluctuations~\cite{OM,WIM,M07,WTMF2009,WM_pss}: 
The $T$-linear resistivity and residual resistivity peak have also been shown theoretically 
by the model Eq.~(\ref{eq:PAM})~\cite{holms,MM}. 

\subsection{Ground state and finite-$T$ phase diagrams}

\begin{figure}
\begin{center}
\includegraphics[width=150mm]{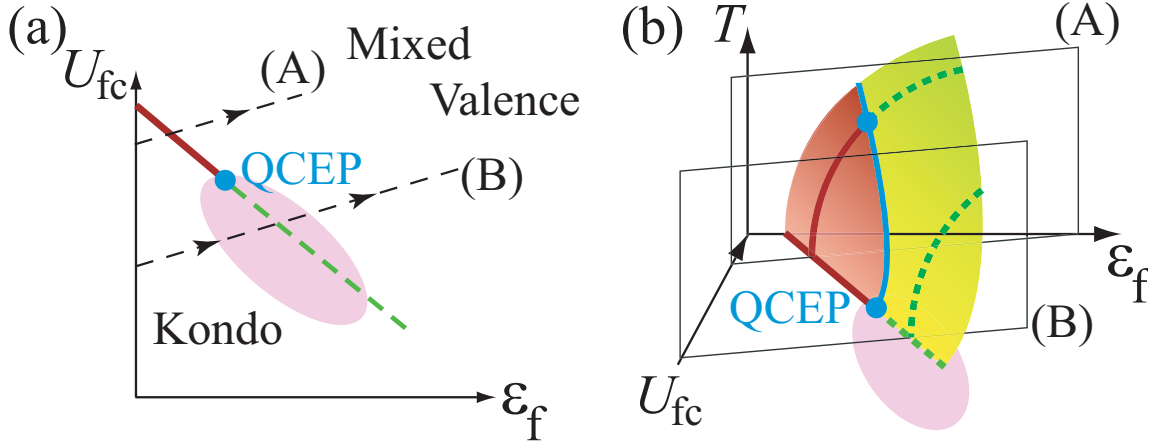}
\end{center}
\caption{\label{fig:PD}(color online) 
Schematic phase diagrams of (a) $\varepsilon_{\rm f}$-$U_{\rm fc}$ plane at $T=0$~K and 
(b) $T$-$\varepsilon_{\rm f}$-$U_{\rm fc}$ space for finite c-f hybridization $V$ and 
large $U$ in the periodic Anderson model Eq.~(\ref{eq:PAM}) (see text). 
The first order valence transition line (bold line) terminates at the quantum critical end point (QCEP) 
(filled circle) and sharp valence crossover occurs on the thick dashed line. 
In (b), the FOVT surface (dark surface) and the valence-crossover surface (light surface) 
are separated by the critical-end line (thick line) touched at $T=0$~K, which is the QCEP. 
In (a) and (b), 
shaded regions at $T=0$~K represent superconducting phases. 
The thin dashed lines with arrow(s) in (a) represent routes for applying pressure to Ce metal (A) and Ce-based 
compounds (B), respectively. Corresponding $T$-$P$ phase diagrams are shown by the cutout (A) and (B) 
in (b), which correspond to Fig.~\ref{fig:PcPv}(a) and Fig.~\ref{fig:PcPv}(b) (or Fig.~\ref{fig:PcPv}(c)), 
respectively (see text). 
}
\end{figure}

Figure~\ref{fig:PD}(a) shows the schematic ground-state phase diagram for paramagnetic states 
drawn on the basis of the DMRG calculation 
in one-spatial dimension $(d=1)$~\cite{WIM} 
and the calculation by the dynamical-mean-field-theory (DMFT) in infinite-spatial dimension 
$(d=\infty)$~\cite{Sugibayashi} applied to the model Eq.~(\ref{eq:PAM}). 
Here,  
the filling $n\equiv (n_{\rm f}+n_{\rm c})/2$ 
with $n_{\rm a}\equiv \sum_{\sigma}\sum_{i=1}^{N}\langle a_{i\sigma}^{\dagger}a_{i\sigma}\rangle/N$ 
for a=f or c with $N$ being the number of lattice sites
is set slightly smaller than 1 (half filling), which describes typical Ce- and Yb-based heavy-fermion metals.

The FOVT line (solid line) terminates at the QCEP (filled circle) 
and sharp valence crossover occurs (dashed line).  
When $\varepsilon_{\rm f}$ is deep enough, the Kondo state with $n_{\rm f}=1$ is realized. 
As $\varepsilon_{\rm f}$ increases across the FOVT and valence-crossover line, 
the mixed valence (MV) state with $n_{\rm f}<1$ is realized~\cite{defMV}. 
At the QCEP, the valence fluctuation 
$\chi_{\rm v}\equiv -\partial n_{\rm f}/\partial\varepsilon_{\rm f}$ 
diverges and at the valence-crossover line, enhanced valence fluctuation appears~\cite{WIM,WTMF2009}. 
We note that 
the slave-boson mean-field theory for $d=1$~\cite{WIM}, $d=2$~\cite{WM2010}, 
and $d=3$~\cite{WTMF2008,WTMF2009} also gives essentially the same phase diagram as Fig.~\ref{fig:PD}(a). 
This is because 
the valence transition and its fluctuation are caused by the atomic Coulomb interaction $U_{\rm fc}$ 
in Eq.~(\ref{eq:PAM}), which is ascribed to the local origin. 
Hence, 
the essential feature of the phase diagram for the valence transition 
does not depend on the spatial dimension. 
As shown by the shaded region in Fig.~\ref{fig:PD}(a), 
near the QCEP of the FOVT, the superconducting phase is shown to appear 
by the slave-boson mean field theory taking into account the Gaussian fluctuations~\cite{OM}, 
which is supported by the DMRG calculation in the same model Eq.~(\ref{eq:PAM}) in $d=1$~\cite{WIM}.

Here, we remark that 
the valence transition is completely different from the localized-to-itinerant transition of f electrons. 
Everywhere in the phase diagram in Fig.~\ref{fig:PD}(a), f electrons are itinerant with finite c-f hybridization 
$\langle f^{\dagger}_{{\bf k}\sigma}c_{{\bf k}\sigma} \rangle\ne 0$. 
Hence, the number of f electrons is always included in the total volume of the Fermi surface, 
giving rise to the large Fermi surface (see also Fig.~\ref{fig:CeRhIn5}(b)). 
This is naturally understood in terms of the Landau-Luttinger sum rule. 
The existence of the QCEP in the ground-state phase diagram ensures that 
adiabatic continuation holds between the Kondo state and the MV state 
by detouring the QCEP of the first-order-transition line. 

Figure~\ref{fig:PD}(b) shows the schematic phase diagram in the $T$-$\varepsilon_{\rm f}$-$U_{\rm fc}$ space. 
The FOVT surface (dark surface) bends to the side of the MV regime for at least large $U_{\rm fc}$ 
since larger entropy coming from f-spin degrees of freedom in the Kondo regime earns 
the larger free-energy gain~\cite{WTMF2009}. 
The FOVT line in Fig.~\ref{fig:PD}(a) is the line at the bottom of the FOVT surface. 
The critical-end line exhibiting diverging valence fluctuations, i.e., $\chi_{\rm v}=\infty$, 
separates the FOVT surface and the valence-crossover surface (light surface), which 
touches the $T=0$ plane at the QCEP. 
We note that even at the valence-crossover surface, valence fluctuations are enhanced~\cite{WTMF2009}, which 
cause remarkable anomalies in physical quantities in Ce- and Yb-based materials, 
as will be shown in \S3 and \S4.

\subsection{Correspondence to Ce- and Yb-based materials}

As shown in Fig.~\ref{fig:PD}(a), the Kondo (MV) state with $n_{\rm f}\approx 1$ 
$(n_{\rm f}<1)$ is realized for deep (shallow) 
$\varepsilon_{\rm f}$ for a fixed $U_{\rm fc}$. 
In Ce systems, 
the $n_{\rm f}=1$ state corresponds to $\rm Ce^{+3.0}$ 
with a $\rm 4f^1$-electron configuration per Ce site. 
In Yb systems, where $\rm Yb^{+3.0}$ has $\rm 4f^{13}$-electron configuration per Yb site, 
the hole picture is useful, so that the $n_{\rm f}=1$ state 
corresponds to $\rm Yb^{+3.0}$ with a $\rm 4f^1$-hole configuration per Yb site.

When we apply pressure to the Ce (Yb) compounds, 
$\varepsilon_{\rm f}$ increases (decreases), 
since negative ions approach the tail of the wavefunction of 4f electrons (holes) at the Ce (Yb) site. 
The hybridization $V$ and inter-orbital Coulomb interaction $U_{\rm fc}$ also increase. 
In the case of Ce metal, 4f and 5d electrons are located at the same Ce site, 
which makes $U_{\rm fc}$ large~\cite{Ce}. 
In Fig.~\ref{fig:PD}(a), the dashed line with an arrow (A) represents the route for 
applying the pressure to Ce metal. 
Hence, the cutout for the large $U_{\rm fc}$ including the $T$ axis labeled by (A) 
in Fig.~\ref{fig:PD}(b) corresponds to the $T$-$P$ phase diagram of Ce metal shown 
in Fig.~\ref{fig:PcPv}(a). 
Namely, the large critical-end temperature $T\sim 600$~K in Ce metal is naturally understood 
from a large $U_{\rm fc}$ due to its on-site origin in Fig.~\ref{fig:PD}(b). 

On the other hand, in the case of Ce (Yb) compounds with conduction electrons 
supplied from elements other than Ce (Yb), 
the Coulomb repulsion between the 4f electron at the Ce (Yb) site and the conduction electron 
$U_{\rm fc}$ is weaker in general because of its inter-site origin. 
Hence, applying pressure to the Ce compounds corresponds to the route represented by 
the dashed line with arrows (B) in Fig.~\ref{fig:PD}(a). 
The cutout for such a moderate $U_{\rm fc}$ labeled by (B) in Fig.~\ref{fig:PD}(b) is considered to 
correspond to the $T$-$P$ phase diagrams of most of the Ce-based compounds. 
Although the magnetically-ordered phase is not shown in Figs.~\ref{fig:PD}(a) and~\ref{fig:PD}(b) for simplicity 
of explanation here, the cutout (B) corresponds to Fig.~\ref{fig:PcPv}(b) (or Fig.~\ref{fig:PcPv}(c)). 
Namely, in CeCu$_2$(Si/Ge)$_2$ systems, the AF phase is located in the Kondo regime 
with a certain interval to the valence-crossover pressure $P_{\rm v}$ 
in Fig.~\ref{fig:PcPv}(b), where 
the valence-crossover temperature $T_{\rm v}^{*}(P)$ covered by the superconducting dome 
corresponds to the cutout of Fig~\ref{fig:PD}(b). 
When we take into account the magnetic order in the phase diagram, 
depending on the strength of the c-f hybridization $V$ (and also $\varepsilon_{\rm f}$ and $U_{\rm fc}$) 
in Eq~(\ref{eq:PAM}), 
the location of the magnetic-paramagnetic phase boundary changes. 
Hence, the relative position of $P_{\rm c}$ and $P_{\rm v}$ changes as shown in Figs.~\ref{fig:PcPv}(b) 
and~\ref{fig:PcPv}(c). 
The interplay of the magnetic order and valence crossover (or transition) 
is quite important in understanding the actual phase diagrams of Ce- and Yb-based systems, 
which will be discussed in \S4. 

Here, we remind the readers of the fact that even at the $\gamma$-$\alpha$ transition in Ce metal, 
the magnitude of the valence jump is about $0.10$~\cite{Wohlleben} 
as it is in the case of YbInCu$_4$~\cite{Matsuda}, as remarked in \S1. 
We note that the magnitude of the valence change at the valence crossover for moderate $U_{\rm fc}$ 
is smaller than that at the FOVT for large $U_{\rm fc}$, which is expected to be in the order of 
$0.01$. 
Actually, such a tiny change of the Yb valence has been observed at 
the valence-crossover temperature 
$T=T_{\rm v}^{*}\sim 40$~K in YbAgCu$_4$~\cite{Sarrao1999}, 
which will be discussed in \S3.3.

\section{Field-induced valence crossover}
\subsection{Theoretical results}

\begin{figure}
\begin{center}
\includegraphics[width=150mm]{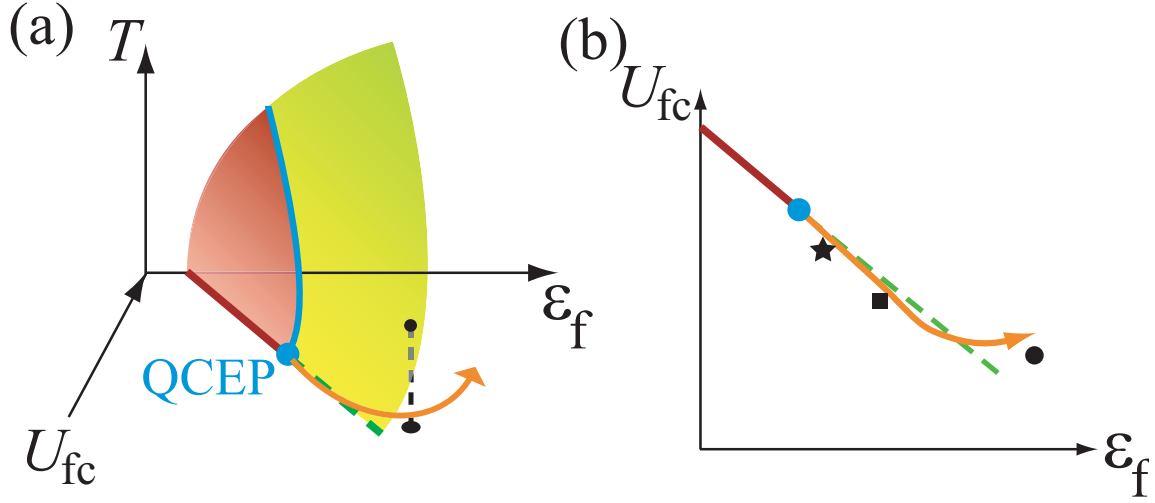}
\end{center}
\caption{\label{fig:PDh}(color online) 
(a) Magnetic-field dependence of FOVT surface and valence-crossover surface 
in the $T$-$\varepsilon_{\rm f}$-$U_{\rm fc}$ phase diagram. 
The FOVT surface (dark surface) and the valence-crossover surface (light surface) 
are separated by the critical-end line (thick line) touched at $T=0$~K, which is the QCEP. 
The thick line with an arrow indicates the locus of the QCEP under magnetic field~\cite{WTMF2008}. 
The black circle indicates a possible location of YbAgCu$_4$ which touches the valence-crossover surface at about $T=40$~K (black dashed line). 
(b) Ground-state phase diagram in $\varepsilon_{\rm f}$-$U_{\rm fc}$ plane. 
The thick line with an arrow represents the locus of the QCEP under a magnetic field. 
The black square and circle indicate possible locations of CeIrIn$_5$ and YbAgCu$_4$, respectively 
(see text and also Ref.~\cite{note_Fig3}). 
The black star indicates a possible location of YbAuCu$_4$. 
}
\end{figure}

As shown in \S2, 
the valence transition is essentially ascribed to the charge degrees of freedom 
and it is nontrivial how the magnetic field affects the valence transition. 
Recently, theoretical studies have clarified that 
the QCEP of the FOVT as well as the sharp valence crossover 
is induced by applying a magnetic field~\cite{WTMF2009}. 
The slave-boson mean-field theory applied to the model Eq.~(\ref{eq:PAM}) in $d=3$ 
with the Zeeman term $-h\sum_{i}(S_{i}^{{\rm f}z}+S_{i}^{{\rm c}z})$ 
$(h\equiv g\mu_{\rm B}H)$ 
has shown that the FOVT surface and the valence-crossover surface move 
as shown in Fig.~\ref{fig:PDh}(a)~\cite{WTMF2008,WTMF2009}. 
On the locus with an arrow illustrated in Fig.~\ref{fig:PDh}(a), 
the magnetic susceptibility $\chi=\partial m/\partial h$ with 
$m\equiv\sum_{i}
\langle
f^{\dagger}_{i\uparrow}f_{i\uparrow}-f^{\dagger}_{i\downarrow}f_{i\downarrow}
+c^{\dagger}_{i\uparrow}c_{i\uparrow}-c^{\dagger}_{i\downarrow}c_{i\downarrow}
\rangle
/N$ 
diverges. 
Namely, metamagnetism occurs. 
The DMRG calculation applied to the same model in $d=1$ has also shown that the 
field-induced extension of the QCEP of the FOVT to the MV regime 
actually occurs, giving rise to the metamagnetisim~\cite{WTMF2008}. 
This finding is quite important and hence 
we put special emphasis on this result: 
{\it 
At the QCEP of the FOVT, not only the valence fluctuation, i.e., charge fluctuation, 
but also the magnetic susceptibility diverges.
} 

Furthermore, 
this result indicates that even in the intermediate-valence materials, 
which does not show any valence transition at $H=0$, 
the QCEP of the FOVT is induced 
by applying the magnetic field $H$. 
An important point here is that 
as shown in Fig.~\ref{fig:PD}(b), most of the Ce- and Yb-based compounds are located 
in the region for a moderate $U_{\rm fc}$, but not in the region for a large $U_{\rm fc}$ 
causing the FOVT. 
Hence, in the Ce- and Yb-based compounds, 
the valence-crossover temperature $T^{*}_{\rm v}(H)$ with strong valence fluctuations 
can emerge in the $T$-$H$ phase diagrams. 
Here, the magnitude of the characteristic field to make $T^{*}_{\rm v}(H)$ emerge depends on 
how close the location of the material 
is to the QCEP of the FOVT in the phase diagram. 
When the material is located close to the QCEP, metamagnetism as well as non-Fermi liquid 
behavior is expected to be observed by applying even a small magnetic field. 
In the following sections of \S3.2, \S3.3, and \S3.4, 
we discuss the possible relevance of this mechanism to experimental observations.

\subsection{Metamagnesim and non-Fermi liquid behavior in CeIrIn$_5$}

\begin{figure}
\begin{center}
\includegraphics[width=75mm]{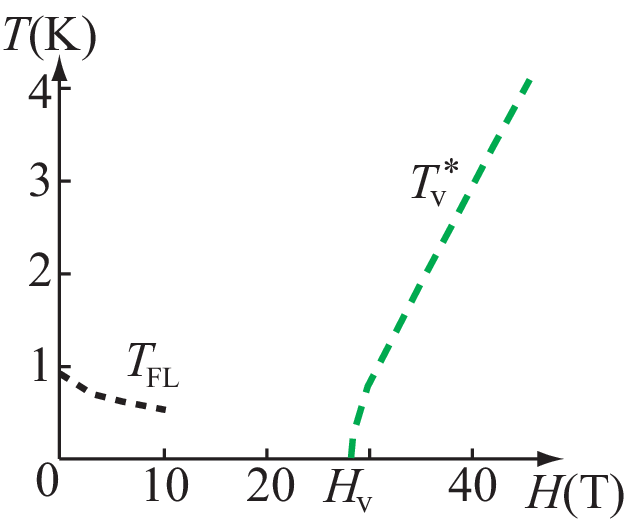} 
\end{center}
\caption{\label{fig:CeIrIn5}(color online) 
$T$-$H$ phase diagram in CeIrIn$_5$~\cite{Capan2009}. 
For $T<T_{\rm FL}$, Fermi-liquid behavior appears in specific heat data~\cite{Capan2004}. 
}
\end{figure}

CeIrIn$_5$ is a heavy-fermion metal, which shows a superconducting transition 
at $T=0.4$~K at ambient pressure~\cite{Petrovic2001}. 
When pressure is applied, the superconducting transition temperature increases 
in spite of the fact that the In-NQR relaxation rate $(T_1T)^{-1}$ drastically decreases, 
suggesting the possibility that a pairing mechanism different from the spin-fluctuation-mediated one 
is relevant to CeIrIn$_5$~\cite{Kawasaki2005}. 
Interestingly, a metamagnetic anomaly was found to exist 
in the magnetization curve in CeIrIn$_5$~\cite{Takeuchi}. 
Several experimental groups have reported this anomaly occurring 
at the crossover (or first-order-transition) line in the $T$-$H$ phase diagram, 
as shown in Fig.~\ref{fig:CeIrIn5}~\cite{Kim,Parm,Capan2004,Capan2009}. 
As $H$ increases to approach $H_{\rm v}$, non Fermi-liquid behavior becomes prominent: 
(i) Convex behavior appears in the $T^{1.5}$ plot of the low-$T$ resistivity suggesting that 
the $T$-linear resisivity appears near $H=H_{\rm v}$~\cite{Capan2004}. 
(ii) Residual resistivity has a peak at $H=H_{\rm v}$~\cite{Capan2009}. 
(iii) The specific-heat constant shows a logarithmic divergence $C/T\sim -\log{T}$ near $H=H_{\rm v}$~\cite{Capan2004}. 
The properties of (i) and (ii) are consistent with the theoretical results of critical valence fluctuations 
in Ref.~\cite{holms} and Ref.~\cite{MM}, 
respectively, as explained in \S1. 
The property of (iii) is also consistent with a theory recently developed for quantum valence criticality, 
which has shown that evaluation of the quasiparticle self-energy 
for one valence-fluctuation exchange process gives the low-$T$ specific heat as 
$C/T\sim -\log{T}$ in a certain-$T$ regime~\cite{WM2010PRL}. 
In Fig.~\ref{fig:CeIrIn5}, the crossover line $T_{\rm v}^{*}(H)$ touches $T=0$~K 
at $H=H_{\rm v}\sim 28$~T~\cite{Capan2009}. 
This fact and the above observations of (i)-(iii) suggest that 
CeIrIn$_5$ is located at the valence-crossover regime indicated by the black square in Fig.~\ref{fig:PDh}(b). 
At $H=0$, the black-square point is located at the Kondo regime. However, as $H$ increases, the slope of the valence crossover line (dashed line) becomes steeper, which crosses the black-square point at $H\sim 28$~T. This causes emergence of the valence-crossover line $T_{\rm v}^{*}(H)$ for $H\ge 28$~T, as shown in Fig.~\ref{fig:CeIrIn5}. 
Namely, CeIrIn$_5$ seems to be located at a distance of about $H=H_{\rm v}\sim 28$~T from 
the valence-crossover line (dashed line in Fig.~\ref{fig:PDh}(b)), close to the QCEP at $H=0$.  
Hence, $H=H_{\rm v}$ at $T=0$~K in Fig.~\ref{fig:CeIrIn5} seems to be in close proximity to the QCEP, 
which affects the low-$T$ physics in CeIrIn$_5$.

Recent measurement of the de Haas-van Alphen (dHvA) effect has confirmed that 
the dHvA frequencies are not substantially altered at $H=H_{\rm v}$ in Fig.~\ref{fig:CeIrIn5}~\cite{Capan2009}. 
This implies that the Fermi-surface topology remains essentially the same between 
the $H<H_{\rm v}$ regime and the $H>H_{\rm v}$ regime. 
This is also consistent with the metamagnetism caused by the field-induced QCEP of the FOVT: 
As explained in detail in Fig.~\ref{fig:PD}(a) in \S2.2, the Fermi-surface volume basically 
does not change at the FOVT and valence crossover 
so long as a system remains in a paramagnetic state 
(Of course, in case of the FOVT (valence crossover), the lattice constant 
shows a discontinuous (continuous) change, which gives a change of the wave number itself). 
We also note that no evidence of the folded Brillouin Zone below and above $H_{\rm v}$ 
has been obtained by the dHvA measurement in CeIrIn$_5$~\cite{Capan2009}. 
This result indicates that the field-induced AF-ordered phase is unlikely to be realized at $H_{\rm v}$, 
although the possibility of the magnetic-breakdown effect in the dHvA measurement 
should be carefully examined. 

We also note that CeIrIn$_5$ and CeCoIn$_5$ have almost the same Fermi surfaces according to 
the band-structure calculations~\cite{CeIrIn5_band} and the dHvA measurements~\cite{CeIrIn5_dHvA}. 
Both materials have a quite similar level scheme of the crystalline electric field (CEF)~\cite{CEF}. 
However, in CeCoIn$_5$,  the crossover line (or first-order transition line) 
accompanied by the non-Fermi liquid behavior 
as shown in Fig.~\ref{fig:CeIrIn5} does not appear in the $T$-$H$ phase diagram. 
We note here that by applying pressure to CeCoIn$_5$, residual resistivity and the $T^2$ coefficient $A$ of the low-$T$ resistivity drop drastically~\cite{Sidorov2002}. These observations suggest that the QCEP of the FOVT, or the sharp valence crossover point exists on a  slightly-negative pressure side in the $T$-$P$ phase diagram of CeCoIn$_5$. Actually, the experimental fact of the emergence of the $T$-linear resistivity and logarithmic divergence of the specific heat $C/T\sim -\log{T}$ at low temperatures and ambient pressure in CeCoIn$_5$ does not contradict this point of view~\cite{WM2010PRL}.

These results indicate that a viewpoint of the closeness to the QCEP of the FOVT is important 
for understanding the Ce115 systems in addition to the conventional view based on 
the competition between the Kondo effect and the RKKY interaction. 
Indeed, this viewpoint offers us a key to resolving the outstanding puzzle in CeRhIn$_5$ as well, 
which will be discussed in \S4.1. 
Since the superconducting correlation has been shown to develop near the QCEP of the FOVT 
theoretically (see Figs.~\ref{fig:PD}(a) and \ref{fig:PDh}(b))~\cite{OM,WIM}, the proximity to 
the QCEP seems to control the occurrence of the unconventional superconductivity in CeIrIn$_5$. 

To examine our theoretical proposal, it is desirable to measure the valence change of Ce 
at the crossover line $T_{\rm v}^{*}$ (or first-order transition line) in Fig.~\ref{fig:CeIrIn5}. 
The X-ray adsorption spectra for direct observation of the Ce valence, the NQR measurement 
for the electric-field gradient, the ultrasonic measurement, and the X-ray diffraction and/or 
thermal expansion measurement for the lattice-constant change are highly desirable.

\subsection{Enhanced magnetic susceptibility and metamagnetism in YbAgCu$_4$}

As noted in \S1, 
YbInCu$_4$ shows the FOVT at $T=T_{\rm v}=42$~K~\cite{Felner,Sarrao1996}. 
When In is replaced by other elements, the FOVT has not been observed. 
However, anomalous behavior which seems to be related to valence fluctuations has been 
observed in YbAgCu$_4$~\cite{Sarrao1999}. 
In YbAgCu$_4$, uniform magnetic susceptibility has a broad peak at $T=40$~K. 
Below $T=40$~K, 
the volume expansion occurs~\cite{Koyama} 
simultaneously with a sharp valence crossover from Yb$^{+2.89}$ for $T>40$~K to 
Yb$^{+2.87}$ for $T<40$~K~\cite{Sarrao1999}, 
indicating that negative volume expansion occurs as $T$ increases to approach $T=40$~K. 
Namely, the uniform magnetic susceptibility is enhanced 
at the valence-crossover temperature $T=T_{\rm v}^{*}=40$~K with strong valence fluctuations. 
This is consistent with our theoretical result discussed in \S3.1~\cite{WTMF2009}. 
At the QCEP and the critical-end line in Fig.~\ref{fig:PDh}(a), 
the valence fluctuations diverge. At the same time, the magnetic susceptibility diverges~\cite{WTMF2008}. 
YbAgCu$_4$ seems to be located in the valence-crossover regime indicated by the black circle in Fig.~\ref{fig:PDh}(a), 
whose distance from the valence-crossover surface at $H=0$ is about $T=40$~K 
as shown by the black dashed line in Fig.~\ref{fig:PDh}(a)~\cite{WTMF2009,WM_pss}. 
Hence, the broad peak of the magnetic susceptibility is considered to be caused by 
the valence fluctuations developed at the valence-crossover surface. 
YbAgCu$_4$ shows a metamagnetism in the low temperature limit around $H=40$~T~\cite{Sarrao1999}. 
This can also be naturally understood from the field-induced valence-crossover surface 
shown in Fig.~\ref{fig:PDh}(a). 
Namely, the location of YbAgCu$_4$ indicated by the black circle in Fig.~\ref{fig:PDh}(b) 
is at a distance of about $H\sim 40$~T from the valence-crossover line close to the QCEP at $H=0$. 
For more details, the readers can refer to Refs.~\cite{WM_pss} and 
\cite{WTMF2009}. 

\subsection{Enhanced magnetic susceptibility and metamagnetism in YbIr$_2$Zn$_{20}$}

Recently, behaviors similar to those of YbAgCu$_4$ have been observed in YbIr$_2$Zn$_{20}$~\cite{Takeuchi2010}. 
The uniform magnetic susceptibility has a peak at $T=T_{\rm v}^{*}=7.4$~K accompanied by 
the volume expansion for $T<T_{\rm v}^{*}$. 
Since the unit-cell volume with the Yb$^{+2.0}$ ($\rm 4f^0$ hole configuration) state 
is larger than that with the Yb$^{+3.0}$ ($\rm 4f^1$ hole configuration) state, 
this observation suggests that the enhancement of the magnetic susceptibility is caused by 
the enhanced Yb-valence fluctuations associated with the valence change of Yb. 
This viewpoint is consistent with the fact that the thermal expansion coefficient is negative 
for $T<T_{\rm v}^{*}$~\cite{Takeuchi2010}. 

Recently, it has been found that 
YbIr$_2$Zn$_{20}$ shows a metamagnetism at $H\sim H_{\rm m}=10$~T~\cite{Takeuchi2010}. 
Near the metamagnetic field $H_{\rm m}$, enhancement of the cyclotron mass of electrons 
has been observed by the dHvA measurement. 
The specific-heat constant $C/T$ and the $T^2$ coefficient of the resistivity at low temperatures 
also indicate the mass enhancement of electrons at $H=H_{\rm m}$. 
Furthermore, residual resistivity has a peak at $H=H_{\rm m}$. 
The temperature region where the low-$T$ resistivity shows the $T^2$ dependence 
becomes narrowest in the vicinity of $H=H_{\rm m}$, which suggests a tendency of 
the $T$-linear resistivity near $H=H_{\rm m}$. 
All these observations are naturally explained by the field-induced valence crossover discussed in \S3.1. 
Although 
the mass enhancement and the residual-resistivity peak suggest a clear signature of 
the Yb-valence fluctuations developed when $H$ approaches $H_{\rm m}$, 
the experimental fact that the resistivity has the $T^2$ dependence at the lowest temperatures even at $H=H_{\rm m}$ 
implies that YbIr$_2$Zn$_{20}$ at ambient pressure is located at the valence-crossover regime. 
Namely, YbIr$_2$Zn$_{20}$ seems to be located at a similar position to that of YbAgCu$_4$, indicated by the black circle in Fig.~\ref{fig:PDh}(b).  The distance from the valence-crossover line close to the QCEP at $H=0$ seems to be about $H\sim 10$~T in YbIr$_2$Zn$_{20}$, since metamagnetic field is about $H_{\rm m}\sim 10$~T~\cite{note_Fig3}. 
Indeed, a steep volume shrinkage has been observed at $H\sim H_{\rm m}$~\cite{Takeuchi2010}, 
which indicates that the Yb valence sharply increases at $H=H_{\rm m}$.
The dHvA measurement has concluded that 
the Fermi surfaces do not change at $H=H_{\rm m}$ 
from the fact that the dHvA frequencies do not change across $H_{\rm m}$~\cite{Takeuchi2010}. 
This is also consistent with the mechanism of the field-induced valence crossover, 
since the Fermi-surface volume is essentially unchanged at the valence crossover 
so long as the system remains in the paramagnetic state 
as discussed in \S2.2 and also in \S3.2. 

When pressure is applied to YbIr$_2$Zn$_{20}$, 
the metamagnetic field $H_{\rm m}$ can be tuned to approach $H=0$~T~\cite{Honda}. 
This seems to be consistent with the location of YbIr$_2$Zn$_{20}$  mentioned above,  since applying pressure makes $U_{\rm fc}$ large because of the reduction of the distance between wavefunctions of 4f and conduction electrons 
and also makes $\varepsilon_{\rm f}<0$ small (i.e., $|\varepsilon_{\rm f}|$ large) in the hole picture. 
Here, we should note that the panel of the cutout shown in Fig.~\ref{fig:PD}(b) is drawn for the Ce compounds. 
In the case of Yb compounds, the panel tilts with a certain angle with the opposite sign 
to the $\varepsilon_{\rm f}$ axis. 
The observation of the quantum criticality at the QCEP of the FOVT by tuning the control parameters 
of the pressure and/or magnetic field is an interesting future issue~\cite{Ohya}. 
Systematic measurements in YbX$_2$Zn$_{20}$ (X=Ir, Co, and Rh) 
are highly desirable for a unified understanding of the effect of the critical valence fluctuations.

\section{Interplay of magnetic order and valence fluctuations}

In the preceding sections, we have discussed the nature of the QCEP of the FOVT 
in a paramagnetic phase 
and how it is controlled by 
applying the pressure and/or magnetic field. 
In Ce- and Yb-based heavy fermion systems, however, the magnetic order can also occur. 
The interplay of the magnetic order and Ce- or Yb-valence fluctuations offers a key concept 
for understanding these systems. 
Below we focus on CeRhIn$_5$ as a prototypical material to demonstrate how such an interplay 
plays a crucial role in understanding the puzzling behaviors.

\begin{figure}
\begin{center}
\includegraphics[width=75mm]{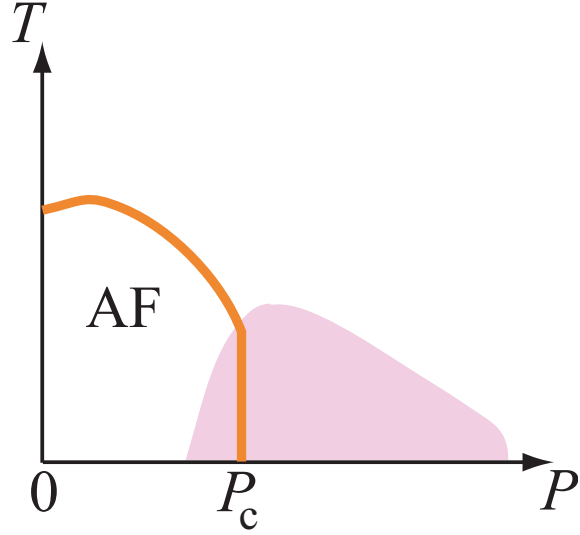}  
\end{center}
\caption{\label{fig:TP_CeRhIn5}(color online) 
Schematic $T$-$P$ phase diagram of CeRhIn$_5$~\cite{Knebel2008,Park2008}. 
Shaded region represents superconducting phase. 
}
\end{figure}

\subsection{CeRhIn$_5$}

CeRhIn$_5$ is a heavy-fermion metal~\cite{Hegger} which undergoes an AF transition 
at $T_{\rm N}=3.8$~K with the ordered vector ${\bf Q}=(1/2, 1/2, 0.297)$ at ambient pressure~\cite{Bao}. 
When pressure is applied, the AF-ordered phase changes to the paramagnetic and superconducting phase 
at $P=P_{\rm c}\sim 2$~GPa, as shown schematically 
in Fig.~\ref{fig:TP_CeRhIn5}~\cite{Knebel2008,Park2008,yashima2009,Shishido2005,Muramatsu,Park2009,note_Pc}. 
This material has attracted much attention since accumulated experiments 
offer outstanding puzzle, whose significant features are summarized as follows: 
(i) The Sommerfeld constant $\gamma_{\rm e}\sim 50$~mJ/(molK$^2$) in the AF state at $P=0$ is about 10-times larger than 
$\gamma_{\rm e}=5.7$~mJ/(molK$^2$) in LaRhIn$_5$~\cite{Hegger,Shishido2005}.
(ii) The Fermi surfaces similar to those in LaRhIn$_5$ in the AF phase 
for $P<P_{\rm c}\sim 2.35$~GPa change to the Fermi surfaces similar to those in CeCoIn$_5$ 
in the paramagnetic phase for $P>P_{\rm c}$ by the dHvA measurement 
performed under the magnetic field $H\sim 15$~T~\cite{Shishido2005}. 
(iii) The cyclotron mass of electrons shows an enhancement toward $P=P_{\rm c}$: 
The cyclotron mass of the $\beta_2$ branch, whose Fermi surface has a cylindrical shape, 
changes from $6m_0$ at $P=0$ to $60m_0$ at $P\lsim P_{\rm c}$ and 
the signal is not detected for $P>P_{\rm c}$, probably because of a too heavy 
mass, $\sim 100m_{0}$~\cite{Shishido2005}. 
(iv) The $T$-linear resistivity emerges prominently near $P=P_{\rm c}$~\cite{Knebel2008,Park2008,Muramatsu}. 
(v) Residual resistivity has a peak at $P\sim P_{\rm c}$~\cite{Knebel2008,Park2008,Muramatsu}. 
(vi) The superconducting phase exists in a wide pressure region around $P=P_{\rm c}$~\cite{Knebel2008,Park2008,yashima2009}.

Because of the dHvA measurement noted above (ii), one might succumb to the temptation to believe 
a scenario in which  
the localized to itinerant transition of f electrons happens at $P=P_{\rm c}$~\cite{Si,Coleman}. 
However, this scenario encounters a serious difficulty in explaining the experimental fact (i) above: 
The AF state with the 10-times mass enhancement at $P=0$ strongly suggests that 
the heavy quasiparticles contribute to the formation of the AF state, 
indicating the existence of the c-f hybridization even in the AF phase for $P<P_{\rm c}$. 
Hence, the theoretical explanation for resolving {\it a series of} puzzles outlined above (i)-(vi) 
in a natural and unified way has been desired. 

First of all, let us point out that the transport anomalies above (iv) and (v) and 
the robust superconducting phase (vi) are quite similar to the observations 
in the CeCu$_2$(Si/Ge)$_2$ systems~\cite{Jaccard,holms,yuan} introduced in \S1. 
Emergence of the $T$-linear resistivity prominent at $P=P_{\rm v}$ in Fig.~\ref{fig:PcPv}(b), 
at which the residual resistivity has a peak, covered with the superconducting phase in the wide pressure region 
is the common feature in CeCu$_2$Ge$_2$~\cite{Jaccard}, 
CeCu$_2$Si$_2$~\cite{holms}, and 
CeCu$_2$(Ge$_{x}$Si$_{1-x}$)$_2$~\cite{yuan}. 
Since the $T$-linear resistivity and residual resistivity peak appear at 
the AF-paramagnetic boundary in CeRhIn$_5$, 
$P_{\rm c}\approx P_{\rm v}$ seems to be realized in Fig.~\ref{fig:TP_CeRhIn5}. 
As will be shown below, 
this is actually the case realized in the model Eq.~(\ref{eq:PAM}) 
for realistic parameters for CeRhIn$_5$, 
which naturally resolves the above puzzles (i)-(vi)~\cite{WM2010}. 

\begin{figure}
\begin{center}
\includegraphics[width=150mm]{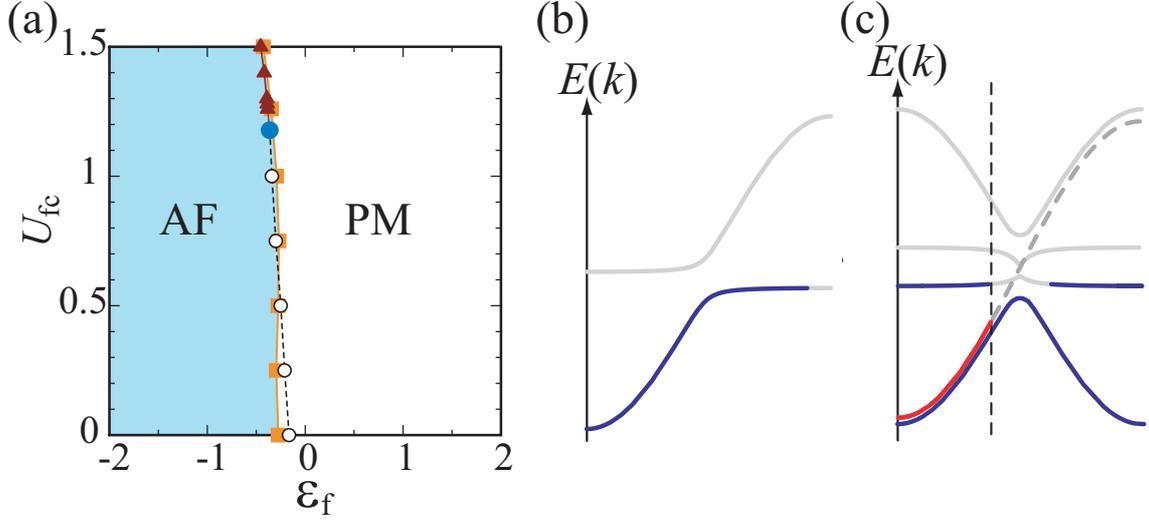}
\end{center}
\caption{\label{fig:CeRhIn5}(color online) 
(a) Ground state phase diagram in $\varepsilon_{\rm f}$-$U_{\rm fc}$ plane 
of the model Eq.~(\ref{eq:PAM}) 
for $V=0.2$ and $U=\infty$ at $n=0.9$ (see text)~\cite{WM2010}. 
The FOVT (solid line with triangles) terminates at QCEP (filled circle). 
Valence crossover with enhanced valence fluctuations occurs at the dashed line 
with open circles. The solid line with filled squares represents 
the antiferromagnetic (AF)-paramagnetic (PM) boundary. 
Occupied (solid lines) and empty (gray lines) 
bands in the periodic Anderson model at $n=(n_{\rm f}+n_{\rm c})/2=0.9$ are shown 
in the paramagnetic phase (b) and in the AF-ordered phase (c). 
In (c), the gray dashed line indicates the energy band of the conduction band, 
$\varepsilon_{\bf k}$ at $n_{\rm c}=0.8$. 
The black dashed line is a guide for the eyes indicating that the Fermi surface in AF-ordered phase 
with finite c-f hybridization coincides with the small Fermi surface where f electrons are 
completely localized. 
}
\end{figure}

Figure~\ref{fig:CeRhIn5}(a) shows the ground-state phase diagram of the model Eq.~(\ref{eq:PAM}) 
without the magnetic field 
on the square lattice 
obtained by the slave-boson mean-field theory~\cite{KR}, which enables us to treat the AF order 
and valence transition or fluctuations on equal footing. 
Here, we set small hybridization $V=0.2$ and the conduction band as 
$\varepsilon_{\bf k}=-2t(\cos k_x+\cos k_y)$ at the total filling $n=0.9$  
to mimic the $\beta_2$ branch of CeRhIn$_5$~\cite{Shishido2005}. 
The FOVT line (solid line with triangles) terminates at the QCEP (filled circle). 
Valence crossover with enhanced valence fluctuations occurs on the dashed line 
with open circles. 
These are the results within the paramagnetic states. 
When the AF order with the ordered vector ${\bf Q}=(\pi,\pi)$ is taken into account, 
we find that the AF-paramagnetic boundary represented by the solid line with filled squares 
emerges at almost the same place as the 
FOVT line and the valence-crossover line. 
Since $\varepsilon_{\rm f}$ can be regarded as the pressure as explained in \S2.3, 
this result implies that $P_{\rm c}\approx P_{\rm v}$ is realized. 

\begin{figure}
\begin{center}
\includegraphics[width=100mm]{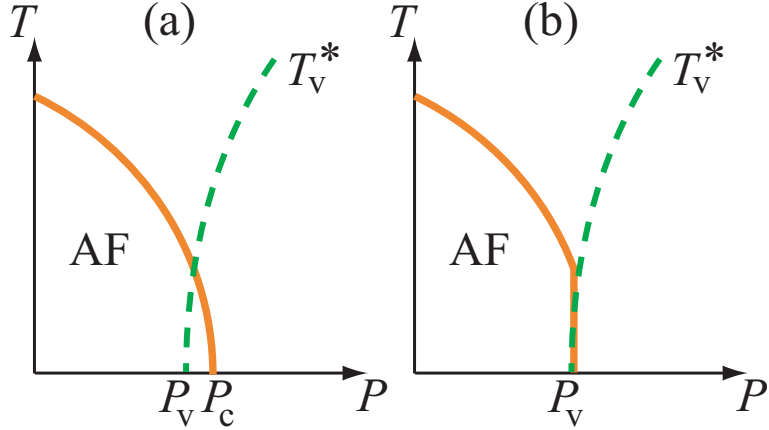}
\end{center}
\caption{\label{fig:PcPv2}(color online) 
(a) Schematic $T$-$P$ phase diagram for $P_{\rm v}<P_{\rm c}$. 
(b) Enhanced valence fluctuations at $P_{\rm v}$ suppress the AF order, giving rise to the AF transition of the first order at $P=P_{\rm v}$ shown by vertical thick line~\cite{WM2010}. 
}
\end{figure}

The precise explanation of this mechanism is illustrated in Figs.~\ref{fig:PcPv2}(a) and~\ref{fig:PcPv2}(b). As shown in Fig.~\ref{fig:PcPv2}(a), when the N${\rm\acute{e}}$el temperature close to $0$~K, i.e., $T_{\rm N}(P)\sim 0$~K, meets the valence-crossover temperature $T_{\rm v}^{*}(P)$, enhanced valence fluctuations suppress the AF order. Then, the AF order is suddenly cut around $P=P_{\rm v}$, as shown in Fig.~\ref{fig:PcPv2}(b). Actually, the slave-boson mean-field calculation shows that the AF order exhibits the first-order transition at the solid line with filled squares in Fig.~\ref{fig:CeRhIn5}(a)~\cite{WM2010}. Namely, the coincidence of the AF transition and the valence crossover or transition is caused by the enhanced valence fluctuations or the FOVT. Hence, the $P_{\rm c}\approx P_{\rm v}$ illustrated in Fig.~\ref{fig:PcPv}(c), or more precisely in Fig.~\ref{fig:PcPv2}(b), is shown to be realized for a small hybridization case in the model Eq.~(\ref{eq:PAM}), which is a realistic parameter for CeRhIn$_5$~\cite{Shishido2005}.
We should remark that for larger hybridization $V$, the AF-paramagnetic boundary appears at smaller $\varepsilon_{\rm f}$ than the FOVT and valence-crossover line in Fig.~\ref{fig:CeRhIn5}(a), which corresponds to Fig.~\ref{fig:PcPv}(b). Hence, depending on the strength of the hybridization $V$, the relative position of $P_{\rm c}$ and $P_{\rm v}$ changes as shown in Figs.~\ref{fig:PcPv}(b) and~\ref{fig:PcPv}(c), and Fig.~\ref{fig:PcPv2}(a).

We have also shown that the Fermi surface calculated for $V=0.2$ in the AF phase at $n=(n_{\rm f}+n_{\rm c})/2=0.9$ 
is nearly the same as the small Fermi surface without c-f hybridization where electrons for $n_{\rm f}=1$ are located 
at the localized f level and  the conduction band is filled up with all extra electrons for $n_{\rm c}=0.8$~\cite{WM2010}. 
This naturally explains the dHvA measurement (ii) above. 
In the paramagnetic phase in Fig.~\ref{fig:CeRhIn5}(a), the large Fermi surface, which counts the f-electron number 
in the total Fermi volume, i.e., $n_{\rm f}+n_{\rm c}$, is realized, as shown in Fig.~\ref{fig:CeRhIn5}(b). 
In the AF phase, the lower and upper hybridized bands are folded, as shown in Fig.~\ref{fig:CeRhIn5}(c). 
Since the lower folded hybridized band is completely filled, the Fermi surface becomes the same as that of 
the conduction electrons at the filling $n_{\rm c}$. 
Hence, we stress that {\it the small Fermi surface appears in the AF phase with finite c-f hybridization}, 
$\langle f^{\dagger}_{{\bf k}\sigma}c_{{\bf k}\sigma}\rangle\ne 0$. 

The mass enhancement observed by the dHvA measurement (iii) above is also quantitatively 
reproduced by the same model Eq.~(\ref{eq:PAM}) under the magnetic field $H=15$~T~\cite{WM2010}. Namely, 
the mass enhancement of the two-dimensional-like Fermi surface of the $\beta_2$ branch 
from $6m_{0}$ at $P=0$ to $60m_{0}$ at $P\lsim P_{\rm c}$~\cite{Shishido2005} is 
well reproduced by the present mechanism, which shows that 
the total density of states at the Fermi level increases near the AF-paramagnetic boundary in Fig.~\ref{fig:CeRhIn5}(a). 
This is because when the pressure is applied to the AF state, 
i.e., when $\varepsilon_{\rm f}$ increases, 
the gap between the original lower hybridized band and the folded band increases as in Fig.~\ref{fig:CeRhIn5}(c). 
Then, the f-electron-dominant flat part of the folded lower hybridized band approaches the Fermi level, 
giving rise to the increase in the density of states. 
In the paramagnetic phase, as $\varepsilon_{\rm f}$ approaches the AF-paramagnetic boundary in Fig.~\ref{fig:CeRhIn5}(a), 
i.e., as $\varepsilon_{\rm f}$ decreases, $n_{\rm f}$ increases to approach 1, which is in the so called Kondo regime. 
Hence, the density of states at the Fermi level increases. 
Thus, the reason why the mass enhancement occurs toward the AF-paramagnetic boundary $P=P_{\rm c}$ in CeRhIn$_5$ 
is naturally explained.

This result is quite consistent with a recent experiment on the effective mass of electrons~\cite{Knebel2008}: 
Knebel {\it et al} applied a magnetic field to CeRhIn$_5$, whose magnitude is the same 
as $H=15$~T used in the dHvA measurement~\cite{Shishido2005}. 
After the disappearance of the superconductivity, the $T^2$ dependence of the resistivity appears 
at lowest temperatures. 
Knebel {\it et al} found that the $\sqrt{A}$ with $A$ being the coefficient of the $T^2$ term 
scales nicely with the cyclotron mass of the $\beta_2$ branch under pressure for 
$0\le P \le P_{\rm c}$, i.e., $\sqrt{A}/m^*=$const. scaling holds~\cite{Knebel2008}. 
This scaling indicates that the mass enhancement toward $P=P_{\rm c}$ is essentially caused by 
the effect of the energy band dispersion of quasiparticles~\cite{MMV}. 
We also note that about 10-times mass enhancement at $P=0$ is also reproduced 
by the same model Eq.~(\ref{eq:PAM})~\cite{WM2010}, which is quite consistent with the experimental fact (i) above.  
Formation of the heavy quasiparticles via the c-f hybridization 
naturally explains the mass enhancement even inside of the AF phase. 

As shown in Fig.~\ref{fig:CeRhIn5}(a), $P_{\rm c}\approx P_{\rm v}$ is a consequence of the suppression of 
the AF order by enhanced valence fluctuations for moderate $U_{\rm fc}$, e.g., $U_{\rm fc}\approx 0.5$, 
which is a realistic parameter for CeRhIn$_5$, giving rise to the $T$-linear resistivity and 
the residual resistivity peak at the AF-paramagnetic boundary. 
Hence, the experimental facts (iv) and (v) are naturally explained. 
Since the superconducting phase was shown to be realized around $P_{\rm v}$~\cite{OM,WIM}, 
as shown in Fig.~\ref{fig:PcPv}(b), 
the above (vi) is also consistently explained (see Fig.~\ref{fig:TP_CeRhIn5} and Fig.~\ref{fig:PcPv2}(b)). 
Hence, all of the above experimental observations (i)-(vi) are naturally explained 
in a unified way. 

We note that 
local correlation effects of 4f electrons give rise to the 
large ordered moment of the AF order: Only about $10~\%$ reduction of the full ordered moment for the Kramers doublet 
of the lowest CEF level is realized even near the AF-paramagnetic boundary in Fig.~\ref{fig:CeRhIn5}(a) 
by the slave-boson mean-field theory~\cite{WM2010}, which is consistent with 
the observation in CeRhIn$_5$~\cite{Park2009}. 
Our result shows that 
strong onsite Coulomb repulsion of 4f electrons yields the local-moment-like behavior 
as if f electrons are localized, in spite of the fact that f electrons are always itinerant, i.e., 
$\langle f^{\dagger}_{{\bf k}\sigma}c_{{\bf k}\sigma}\rangle \ne 0$ always holds 
everywhere of the phase diagram shown in Fig.~\ref{fig:CeRhIn5}(a),  
which is nothing but a manifestation of the localized-itinerant duality~\cite{KM1990}. 

As reported in Ref.~\cite{Pham2006}, the experimental comparison between CeRhIn$_5$ and CeCo(In$_{1-x}$Cd$_{x}$)$_5$ shows that CeRhIn$_5$ at $P\gsim P_{\rm c}$ corresponds to CeCoIn$_5$ at $P=0$~\cite{M07}. 
This is quite consistent with our argument in \S3.2: Because $P_{\rm c}\approx P_{\rm v}$ is realized in CeRhIn$_5$, if the above correspondence holds, that implies that on the slightly-negative pressure side in the $T$-$P$ phase diagram of CeCoIn$_5$, the QCEP or sharp Ce-valence crossover point, i.e., $P=P_{\rm v}$, exists, which is the same conclusion stated in \S3.2.

The interplay of the magnetic order and enhanced valence fluctuations plays an important role not only 
in CeRhIn$_5$ but also in other materials. 
One such example is YbAuCu$_4$, which will be discussed in the next section. 

\subsection{YbAuCu$_4$}

\begin{figure}
\begin{center}
\includegraphics[width=70mm]{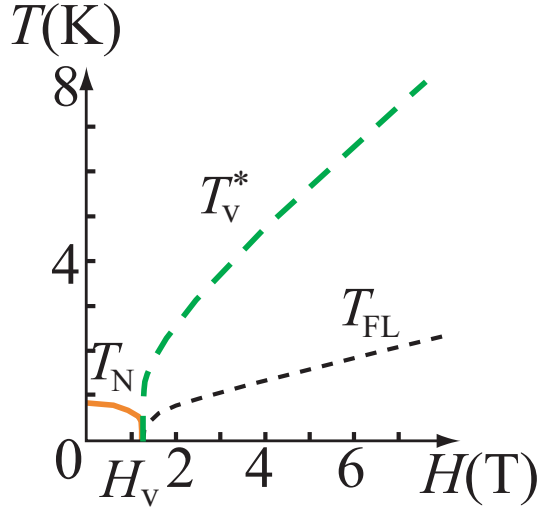}
\end{center}
\caption{\label{fig:YbAuCu4}(color online) 
$T$-$H$ phase diagram of YbAuCu$_4$~\cite{wada}. 
The $\rm N{\acute{e}}el$ temperature $T_{\rm N}$ is suppressed and the Yb valence crossover temperature $T_{\rm v}^{*}$ 
(thick dashed line) 
is induced for $H>H_{\rm v}\sim1.3$~T. 
For $T<T_{\rm FL}$ (black dashed line), the resistivity shows the Fermi liquid behavior $\rho \sim T^2$. 
}
\end{figure}

YbAuCu$_4$ is a sister compound of YbInCu$_4$, which exhibits the FOVT as noted in \S1. 
We note that the phase diagram of YbAuCu$_4$ shown in Fig.~\ref{fig:YbAuCu4} 
is essentially the same as Fig.~\ref{fig:PcPv}(c) or Fig.~\ref{fig:PcPv2}(b), 
if $P$ is replaced by $H$~\cite{yamamoto}. 
The $^{63}$Cu-NQR measurement has detected that there exists a characteristic temperature 
$T_{\rm v}^{*}(H)$ in the $T$-$H$ phase diagram, at which 
the Cu-NQR frequency changes sharply, strongly suggesting that the Yb-valence crossover occurs~\cite{wada}. 
The Yb-valence crossover temperature $T_{\rm v}^{*}(H)$ becomes $T_{\rm v}^{*}=0$~K at the field $H_{\rm v}$, 
which seems to coincide with 
the field $H_{\rm c}$ at which the N$\rm\acute{e}$el temperature goes to $T_{\rm N}=0$~K~\cite{yamamoto,wada} 
(see Fig.~\ref{fig:YbAuCu4}). 
Namely, $H_{\rm c}\approx H_{\rm v}$ occurs in the $T$-$H$ phase diagram. 
As discussed in \S3.1, 
the valence crossover temperature $T_{\rm v}^{*}(H)$ is induced by applying the magnetic field. 
If YbAuCu$_4$ is located at the moderate $U_{\rm fc}$ regime indicated by the black star 
in Fig.~\ref{fig:PDh}(b), 
the appearance of $T_{\rm v}^{*}(H)$ in Fig.~\ref{fig:YbAuCu4} is naturally explained~\cite{WM_pss}. 
Namely, YbAuCu$_4$ seems to be in the Kondo regime at $H=0$ and by applying $H=H_{\rm v}\sim 1.3$~T, 
the valence crossover $T_{\rm v}^{*}(H)$ is induced, as shown in Fig.~\ref{fig:YbAuCu4}. This indicates that YbAuCu$_4$ 
is close to the QCEP at $H=0$ with a distance of about $H\sim 1.3$~T as indicated by the black star in Fig.~\ref{fig:PDh}(b). 
As explained in \S4.1, 
when $T_{\rm N}(H)$ close to $0$~K meets $T_{\rm v}^{*}(H)$, 
a suppression of the magnetic order by enhanced valence fluctuations occurs, 
which gives rise to the coincidence $H_{\rm c}\approx H_{\rm v}$. 
In YbAuCu$_4$, 
the cusp-like anomaly of the residual resistivity was observed at 
$H=H_{\rm c}\approx H_{\rm v}$~\cite{wada}. 
The mass enhancement of electrons estimated from the $A$ coefficient in the resistivity 
was also observed near $H=H_{\rm c}\approx H_{\rm v}$~\cite{wada}. 
These observations are naturally explained by the enhanced Yb-valence fluctuations~\cite{WM_pss}. 

\section{Summary and outlook}

In this article, 
the roles of critical valence fluctuations in Ce- and Yb-based heavy fermion metals are discussed. 
Recent development of theory and experiment has revealed that critical Ce- and Yb-valence fluctuations 
play a key role in several anomalies in this family of materials.   
Magnetic field is shown to be a useful control parameter to induce the quantum critical end point (QCEP) of the 
first-order valence transition (FOVT). 
As a prototypical material, metamagnetism and non-Fermi liquid behavior in 
CeIrIn$_5$, YbAgCu$_4$, and YbIr$_2$Zn$_{20}$ are shown to be naturally explained by the mechanism 
of the field-induced valence crossover.  
The interplay of the magnetic order and Ce- or Yb-valence fluctuations is a key mechanism 
for understanding anomalous behavior in Ce- and Yb-based heavy fermion systems.  
It is shown that the interplay resolves the outstanding puzzle in CeRhIn$_5$. The origin of the 
transport anomalies and drastic change of the Fermi surfaces accompanied 
by the mass enhancement of electrons 
is naturally explained by the interplay in a unified way. 
A special emphasis is put on the fact that a small Fermi surface generally appears by the folding of the 
hybridized band in the AF phase in the periodic Anderson model. 
The unified explanation for CeInIn$_5$ and CeRhIn$_5$ strongly suggests that 
the viewpoint of the closeness to the QCEP of the FOVT is indispensable in understanding the Ce115 systems. 
As another prototypical material where the interplay occurs, 
the $T$-$H$ phase diagram of YbAuCu$_4$ is discussed. 
The proximity to the QCEP of the FOVT is a key concept for understanding Ce- and Yb-based 
heavy fermion systems as an underling mechanism.  

For outlook, 
recently synthesized materials, YbIr$_2$Zn$_{20}$,  
YbCo$_2$Zn$_{20}$, and YbRh$_2$Zn$_{20}$ are promising materials, 
which are expected to show the properties caused by the Yb-valence fluctuations by tuning the control parameters of 
the magnetic field and pressure. 
Recently, in $\beta$-YbAlB$_4$, where the superconductivity was first discovered among Yb-based 
compounds~\cite{nakatuji}, 
evidence of strong Yb-valence fluctuations has been reported~\cite{Okawa}. 
Further study to clarify the role of the valence fluctuations in the unconventional criticality 
as well as the origin of the superconductivity is an interesting future problem.   
We point out that 
the $T$-$H$ phase diagram in YbRh$_2$Si$_2$~\cite{Gegenwart2007},  
where unconventional criticality has attracted much attention 
in correlated electron systems~\cite{Gegenwart2008}, is closely similar to that in YbAuCu$_4$ 
(see Fig.~\ref{fig:YbAuCu4}).  
Hence, there exists a possibility that 
the critical Yb-valence fluctuation plays a key role 
in the origin of the unconventional criticality in YbRh$_2$Si$_2$. 
Thus, it is quite important to examine the possibility of whether the Yb-valence change takes place 
at the crossover temperature $T^{*}(H)$ whose origin has not yet been clarified experimentally 
in the $T$-$H$ phase diagram. 
The experimental fact that YbRh$_2$Si$_2$ shows 
a huge mass enhancement even at $H=0$ 
such as $\gamma_{\rm e}\sim 1.7$~J/(molK$^2$) deep inside of the AF phase~\cite{Krellner2009} 
indicates that heavy quasi particles form the AF-ordered state as in CeRhIn$_5$. 
The Co-NQR measurement in Yb(Rh$_{1-x}$Co$_{x}$)$_2$Si$_2$ and/or the direct observation of 
the Yb valence by 
X-ray adsorption measurement at the crossover temperature $T^{*}(H)$ are highly desirable. 
As noted in \S2.3, the magnitude of the valence change is expected to be of the order of $0.01$ 
in most of the Ce and Yb compounds. 
Hence, high accuracy measurement which can detect the tiny change of the valence 
is a challenging future problem, which will open new avenues of study in this field. 

\section{Acknowledgements}
This work is supported by the Global COE program (G10) and a Grant-in-Aid for Scientific Research (No. 19340099) from the Japan Society for the Promotion of Science (JSPS), and a Grant-in Aid for Scientific Research on Innovative Areas ``Heavy Electrons"@(No.20102008) from the Ministry of Education, Culture, Sports, Science and Technology of Japan. S. W. is supported by a Grant-in-Aid for Young Scientists (B) (No. 21740240) from JSPS.


\begin{thebibliography}{9}
\bibitem{note_Fig3} 
We note that each material has different strength of c-f hybridization $V$. Hence, the locus of the QCEP under the magnetic field depends on $V$ in Figs.~\ref{fig:PDh}(a) and~\ref{fig:PDh}(b), strictly speaking. To avoid confusion by drawing each locus (thick line with an arrow) for each material, we just drew one locus and schematic locations of CeIrIn$_5$, YbAgCu$_4$, and YbAuCu$_4$, respectively, in Fig.~\ref{fig:PDh}(b).
\end{thebibliography}
\end{document}